%% file: SOscattering.tex
\documentclass[aps,amsmath,amssymb,superscriptaddress,longbibliography,preprint]{revtex4-1}

\usepackage{bm}
\usepackage{graphicx}
\usepackage[colorlinks,linkcolor=blue,urlcolor=blue,citecolor=blue,bookmarks=false]{hyperref}
\usepackage{color}

\newcommand{\CEMS}{RIKEN Center for Emergent Matter Science, Wako, Saitama 351-0198, Japan.}
\newcommand{\MSL}{Laboratory for Materials and Structures, Tokyo Institute of Technology, Yokohama, Kanagawa 226-8503, Japan.}

\newcommand{\Gbar}{\overline{\Gamma}}
\newcommand{\Mbar}{\overline{\mathrm M}}
\newcommand{\Kbar}{\overline{\mathrm K}}
\newcommand{\bk}{\bm{k}}
\newcommand{\bkp}{\bk^\prime}
\newcommand{\psin}{\psi_n(\bk)}
\newcommand{\psim}{\psi_m(\bkp)}
\newcommand{\gn}{g_n(\bk,\omega)}
\newcommand{\gm}{g_m(\bkp,\omega)}

\begin{document}

\title{Spin-orbit scattering visualized in quasiparticle interference}

\author{Y. Kohsaka}
\email{kohsaka@riken.jp}
\affiliation{\CEMS}

\author{T. Machida}	\affiliation{\CEMS}
\author{K. Iwaya}	\affiliation{\CEMS}
\author{M. Kanou}	\affiliation{\MSL}
\author{T. Hanaguri}\affiliation{\CEMS}
\author{T. Sasagawa}\affiliation{\MSL}

\date{\today}

\begin{abstract}
In the presence of spin-orbit coupling, electron scattering off impurities depends on both spin and orbital angular momentum of electrons -- spin-orbit scattering.
Although some transport properties are subject to spin-orbit scattering, experimental techniques directly accessible to this effect are limited.
Here we show that a signature of spin-orbit scattering manifests itself in quasiparticle interference (QPI) imaged by spectroscopic-imaging scanning tunneling microscopy.
The experimental data of a polar semiconductor BiTeI are well reproduced by numerical simulations with the $T$-matrix formalism that include not only scalar scattering normally adopted but also spin-orbit scattering stronger than scalar scattering.
To accelerate the simulations, we extend the standard efficient method of QPI calculation for momentum-independent scattering to be applicable even for spin-orbit scattering.
We further identify a selection rule that makes spin-orbit scattering visible in the QPI pattern.
These results demonstrate that spin-orbit scattering can exert predominant influence on QPI patterns and thus suggest that QPI measurement is available to detect spin-orbit scattering.
\end{abstract}

\maketitle

\section{Introduction}

Spin-dependent scattering has played important roles in many fields of physics for a long time.
Spin-dependent asymmetric scattering of electron beams in vacuum provided a foundation of relativistic quantum mechanics~\cite{Mott1929,Shull1943}.
In condensed matter physics, spin-orbit scattering of electrons propagating in solids contributes to some transport phenomena~\cite{Hikami80,Nagaosa10,Sinova15}.
For example, spin-dependent impurity scattering caused by spin-orbit scattering is among the origins of anomalous Hall effect and (extrinsic) spin Hall effect~\cite{Nagaosa10,Sinova15}.
Another direct consequence of spin-orbit scattering is rotation of electron spin, changing interference between wave functions of electrons around an impurity.
This effect on quantum interference is known as the origin of weak anti-localization~\cite{Hikami80}.

Interference of wave functions results in a periodic modulation of the local density of states (LDOS)\@.
This modulation, known as quasiparticle interference (QPI), has been imaged by spectroscopic imaging scanning tunneling microscopy (SI-STM) in a wide variety of materials~\cite{Hasegawa93,Crommie93,Hoffman02,Zhang09,Beidenkopf11,Zeljkovic14,Arguello15,Zheng16,Inoue16}.
QPI has been studied mostly to acquire momentum-resolved information of electronic states from its characteristic periodicity based on an assumption that the scattering center is a scalar one.
This assumption is widely used even for strong spin-orbit coupling systems~\cite{Zhang09,Beidenkopf11,Zeljkovic14,Zheng16,Inoue16} where spin-orbit scattering is also likely to be strong.
However, the role of spin-orbit scattering in QPI is obscure.

A primal difference between spin-orbit and scalar scattering is that the former depends on both spin and momentum of electrons whereas the latter does not.
Consequently, spin-orbit scattering can cause additional enhancement or suppression of QPI that is unanticipated for scalar scattering.
Lee \textit{et al}. theoretically indicate that spin-orbit scattering enhances new scattering channels for the surface states of topological insulator Bi$_2$Te$_3$~\cite{Lee09}.
In the experiments, however, such enhancement and resultant multiple branches of QPI have never been observed~\cite{Zhang09,Beidenkopf11}.

In this paper, exploiting atomic-resolution SI-STM and numerical simulations, we reveal that spin-orbit scattering is predominant for QPI of the quasi-two-dimensional states of a polar semiconductor BiTeI.
We performed a detailed analysis of QPI using the standard $T$-matrix formalism with an extended technique to accelerate calculations including not only momentum-independent scalar scattering but also momentum-dependent spin-orbit scattering.
All the components of QPI observed by the experiments are successfully reproduced when spin-orbit scattering is considered.
The key ingredient is the selection rule due to spin-orbit scattering, which selectively enhances one of the main scattering channels contributing QPI of BiTeI.
Our finding demonstrates that spin-orbit scattering is actually observed in QPI and provides a foothold to get better insight into the electronic states through QPI.

\section{Overview of expermental results of B\MakeLowercase{i}T\MakeLowercase{e}I}

BiTeI has a polar crystal structure with layered stacking of triple layers composed of Te, Bi, and I layers, and hosts giant Rashba-type spin splitting in bulk bands as well as at the surface as observed by angle-resolved photoemission spectroscopy (ARPES)~\cite{Ishizaka11,Crepaldi12,Landolt12,Sakano12,Sakano13}.
A domain structure composed of opposite stacking orders is found in this material, resulting in two kinds of termination at a single surface~\cite{Tournier-Colletta14,Butler14,Fiedler14,Kohsaka15}.
Especially in the Te-terminated areas, differential conductance ($dI/dV$) images show clear QPI of quasi-two-dimensional states split off from the bulk valence band by the spontaneous electric polarization~\cite{Kohsaka15}.

\begin{figure}
	\centering
	\includegraphics{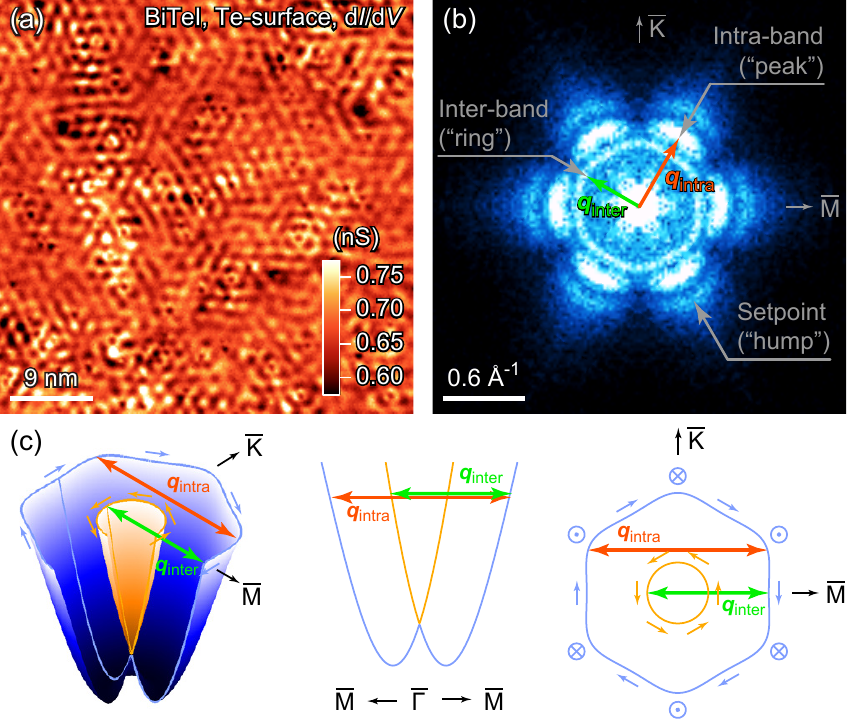}
	\caption{(color online) (a) A 45 $\times$ 45~nm$^2$ $dI/dV$ image taken at the Te-terminated surface of BiTeI\@.
	The image was taken at $-10$~mV with a lock-in modulation voltage of 5~mV$_\mathrm{rms}$ and a setup tunneling current of 0.2~nA at a setup bias voltage of 0.2~V\@.
	(b) Fourier transform of (a).
	(c) Schematic figures of the band structure of quasi-two-dimensional states at the Te-terminated surface observed by ARPES~\cite{Ishizaka11,Crepaldi12,Landolt12}.
	From left to right, a three-dimensional illustration of the band structure, the band dispersion in the $\Gbar$-$\Mbar$ direction, and a constant energy contour.
	The double-headed arrows denote dominant scattering channels producing the QPI.
	Spin directions are depicted by the arrows and markers colored in orange and blue; in-plane components are denoted by the arrows and out-of-plane components are denoted by the markers.}
	\label{fig:exp}
\end{figure}

To define points to be focused, we briefly summarize QPI of BiTeI.
(Experimental details are described in Ref.~\onlinecite{Kohsaka15}.)
Figure~\ref{fig:exp}(a) shows a typical image of QPI in the Te-terminated area.
The Fourier-transformed image shown in Fig.~\ref{fig:exp}(b) reveals that the QPI patterns consist of three major components: the hexagonal ring surrounding the $\Gbar$ point, the strong peaks in the $\Gbar$-$\Mbar$ direction, and the outermost humps in the $\Gbar$-$\Mbar$ direction.
These features are commonly observed in all samples studied.
By comparing QPI dispersions with the ARPES results,~\cite{Ishizaka11,Crepaldi12,Landolt12,Sakano12} the hexagonal ring and the strong peaks are assigned to interband scattering between the spin-split bands and intraband scattering between the corners of the hexagonally warped outer branch, respectively, as depicted in Fig.~\ref{fig:exp}(c).
The outermost humps change their locations depending on bias voltages for stabilizing the scanning tip, and therefore are attributed to the so-called setpoint effect discussed later.
The near-$\Gbar$ feature, which varies from one sample to another, originates from nanometer-scale inhomogeneity due to random distribution of defects.

Although the positions of the hexagonal ring and the strong $\Gbar$-$\Mbar$ peaks are understood as described above, there remains a puzzle in their intensities.
The intraband scattering is mostly forbidden because the backscattering from $\bk$ to $-\bk$ is suppressed due to the anti-parallel spin orientations.
The $\Gbar$-$\Mbar$ peaks are nevertheless allowed because of deviation from the backscattering.
Meanwhile, the interband scattering giving the hexagonal ring is always allowed because the spin orientations are almost parallel.
That is, the spin texture of the band structure is more beneficial for the hexagonal ring than for the $\Gbar$-$\Mbar$ peaks, although the former is actually weaker than the latter.
This inverted intensity is a robust signature of spin-orbit scattering, as revealed below.

\section{The model and $T$-matrix formalism}

\begin{figure}
	\centering
	\includegraphics{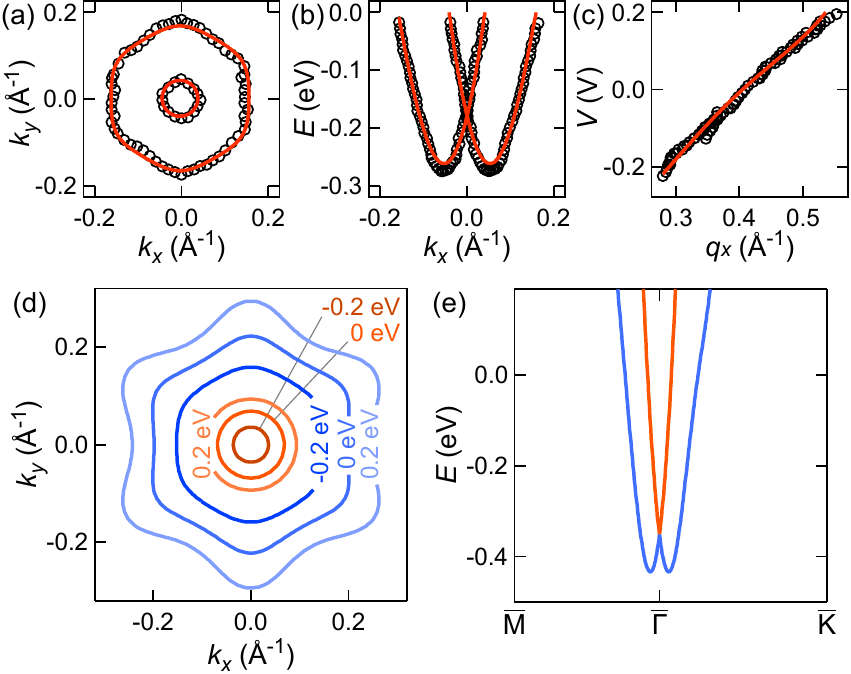}
	\caption{(color online) 
	Fitting results to determine the parameters of Eq.~\eqref{eqn:H0}:
	(a) the Fermi surface,~\cite{Sakano13} (b) band dispersion in the $\Gbar$-$\Mbar$ direction,~\cite{Sakano13} and (c) QPI dispersion the in the $\Gbar$-$\Mbar$ direction~\cite{Kohsaka15}.
	Open circles and solid curves are experimental data and fitting results, respectively.
	Since there is an energy offset $\Delta E_0$ between ARPES and QPI dispersions, $E_0$ is given by the sum of two fitting parameters, $E_0=E_0^\mathrm{ARPES}+\Delta E_0$, where $E_0^\mathrm{ARPES}=-0.179$~eV and $\Delta E_0=-0.173$~eV.
	The eigenvalues of the model Hamiltonian are shown in (d) constant energy contours and (e) dispersion along $\Mbar$-$\Gbar$-$\Kbar$.}
	\label{fig:fit}
\end{figure}

To solve the puzzle of intensity, we numerically simulate QPI patterns.
To model the quasi-two-dimensional state originating from the bulk valence band predominated by Bi 6$p_z$ orbitals,~\cite{Bahramy11} we employ an extended Rashba Hamiltonian,
\begin{align}
	H_0(k_x,k_y) &= \left(E_0 + \frac{k^2}{2m}E(k)\right)I + V(k)(k_x\sigma_y-k_y\sigma_x) + \Lambda(k)(3k_x^2-k_y^2)k_y\sigma_z,
	\label{eqn:H0}
\end{align}
where $I$ and $\sigma_i$ $(i = x, y, z)$ are the identity matrix and the Pauli matrices, respectively, with $k=\sqrt{k_x^2+k_y^2}$.
If $E(k)$ and $V(k)$ are constant and $\Lambda(k)=0$, Eq.~\eqref{eqn:H0} gives the Bychkov-Rashba Hamiltonian~\cite{Bychkov84}.
The last term of Eq.~\eqref{eqn:H0} reflects C$_\mathrm{3v}$ symmetry of BiTeI~\cite{Ishizaka11,Bahramy12}.
We extend $H_0$ up to $k^6$ terms, $E(k)=1+\alpha_4k^2 + \alpha_6k^4$, $V(k)=v(1+\beta_3k^2+\beta_5k^4)$, and $\Lambda(k)=\lambda(1+\gamma_5k^2)$ so that it is invariant under a three-fold rotation along the $z$ direction, mirror operation about the $xz$ plane ($x$ is along the $\Gbar$-$\Mbar$ direction), and the time-reversal operation.
The higher terms up to $k^6$ are required for QPI calculations performed in the whole surface Brillouin zone whereas $k^3$ terms are enough to reproduce the ARPES results near the $\Gbar$ point~\cite{Sakano13}.
We choose parameters as
$m = 0.0168$~eV$^{-1}$\AA$^{-2}$,
$\alpha_4 = -2.03$~\AA$^{-2}$,
$\alpha_6 = 87.5$~\AA$^{-4}$,
$v = 3.13$~eV\AA$^{-1}$,
$\beta_3 = -2.01$~\AA$^{-2}$,
$\beta_5 = 323$~\AA$^{-4}$,
$\lambda = -41.7$~eV\AA$^{-3}$,
$\gamma_5 = 2.43$~\AA$^{-2}$, and
$E_0 = -0.352$~eV
by fitting experimental data (Fig.~\ref{fig:fit}).

QPI patterns have been calculated with the standard $T$-matrix formalism for a single local impurity.
In fact, there are many defects in the field of view of Fig.~\ref{fig:exp}(a).
However, the three major features of QPI are independent of details of defect distribution as evidenced by the experimental fact that they are observed in all samples.
Therefore, we postulate that multiple impurities work on overall intensity in a statistical manner~\cite{Arguello15} and a single impurity is a good starting point to discuss QPI patterns.
The LDOS is written by the retarded Green's function $\hat{G}$ in momentum space,
\begin{align}
	\rho(\bm{q},\omega) = -\frac{1}{2\pi i}\sum_{\bk}\mathrm{Tr}\left\{\hat{G}(\bk,\bk-\bm{q},\omega)-\hat{G}^*(\bk,\bk+\bm{q},\omega)\right\},
	\label{eqn:rhoQ}
\end{align}
where $\rho(\bm{q},\omega)$ is the Fourier transform of the LDOS, $\rho(\bm{q},\omega)=\int\rho(\bm{r},\omega)e^{-i\bm{q}\cdot\bm{r}}d\bm{r}$.
Here we consider the Green's function in matrix form to include spin.
In the presence of a scattering center the potential of which in momentum space is $\hat{V}_{\bk,\bkp}$,
\begin{align}
	\hat{G}(\bk,\bkp,\omega) = \hat{G}_0(\bk,\omega)\delta_{\bk,\bkp} + \hat{G}_0(\bk,\omega)\hat{T}_{\bk,\bkp}(\omega)\hat{G}_0(\bkp,\omega),
	\label{eqn:G}
\end{align}
where the $T$ matrix satisfies
\begin{align}
	\hat{T}_{\bk,\bkp}(\omega)
	= \hat{V}_{\bk,\bkp} + \sum_{\bm p}\hat{V}_{\bk,\bm{p}}\hat{G}_0(\bm{p},\omega)\hat{T}_{\bm{p},\bkp}(\omega).
	\label{eqn:Tseries}
\end{align}
Here $\hat{G}_0$ is the bare Green's function, $\hat{G}_0(\bk,\omega) = \sum_n\gn\big|\psin\big\rangle\big\langle\psin\big|$,  $\gn=\left(\omega+i\eta-\epsilon_n(\bk)\right)^{-1}$, where $\epsilon_n(\bk)$ and $\psin$ are the $n$th eigenvalue and the $n$th eigenstate of the bare Hamiltonian, respectively, with $\eta$ being a small broadening factor (10~meV for our simulations).

QPI patterns can be computed in principle with these equations.
For momentum-independent scattering (e.g., scalar scattering), a direct calculation of the $\bk$ summation requires $\mathcal{O}(N^4)$ operations for a single QPI image, where $N\times N$ is the number of grid points.
The amount of this calculation can be reduced to $\mathcal{O}(N^2\log_2N)$ by using fast Fourier transform (FFT)~\cite{Arguello15,Derry15}.
For momentum-dependent scattering (e.g., spin-orbit scattering), however, the FFT-based technique has not been applied and consequently the $\bk$ summation requiring $\mathcal{O}(N^6)$ operations has been directly calculated.
The enormous amount of calculation has hindered precise and comprehensive analysis of QPI;
calculations have often been done only in a narrow range and at a low resolution of energy and momentum~\cite{Lee09}.
We find that the FFT-based technique is still available for momentum-dependent scattering satisfying a certain condition.
Because of this method, the amount of calculation for spin-orbit scattering can be reduced to $\mathcal{O}(N^2\log_2N)$ that greatly accelerates our simulations.
Details of the method are described in Appendix~\ref{appendix:fastCalculation}.

\section{Numerical results}

\begin{figure}
	\centering
	\includegraphics{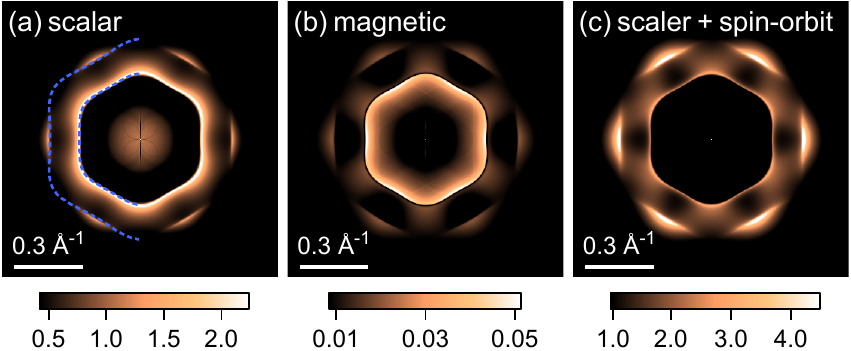}
	\caption{(color online) Fourier-transformed images of QPI patterns at $-10$~meV, $\big|\rho(\bm{q},\omega=-10~\mathrm{meV})\big|$, calculated with different scattering centers.
	(a) Scalar scattering.
	(b) Magnetic scattering.
	(c) Scalar and spin-orbit scattering with $c = 60$~\AA$^2$.
	The blue dashed lines in (a) depict scattering vectors expected from constant energy contours of the band dispersion.
	The inner and outer lines denote interband scattering between the spin-split bands and intraband scattering of the outer branch, respectively.}
	\label{fig:calc}
\end{figure}

We begin our simulations with a scalar impurity.
In this case, the scattering potential is independent of momentum, $\hat{V} = V_0I$, where $V_0$ is strength of scattering ($V_0 = 0.1$~eV for all simulations).
The $T$ matrix is also simplified to a momentum-independent form.
This simplification makes the calculation greatly easy and is why a scalar impurity is widely assumed as scattering center.
The simulation result successfully reproduces the hexagonal ring and the $\Gbar$-$\Mbar$ peaks as shown in Fig.~\ref{fig:calc}(a).
The hexagonal ring appears at the location of interband scattering between the spin-split bands and the $\Gbar$-$\Mbar$ peaks lie slightly outside of intraband scattering of the outer branch, corroborating the peak assignment described above.
No prominent feature appears near the $\Gbar$ point, being consistent with the experiments. (See Appendix~\ref{appendix:JDOS} for details.)
Although these basic features are reproduced, the ring is stronger than the peaks, replicating the puzzle.
This discrepancy in intensities is robust as long as the band parameters are in a reasonable range, suggesting that the scattering is not a simple scalar one.

We then consider two kinds of scattering, magnetic scattering and spin-orbit scattering.
Since both rotate electron spin, they may change the situation that is beneficial for the hexagonal ring.
Figure~\ref{fig:calc}(b) shows the calculation result with magnetic scattering $\hat{V}=V_0\sigma_z$ corresponding to a classical magnetic moment pointing in the $z$ direction.
The $\Gbar$-$\Mbar$ peaks are still weaker than the hexagonal ring and the QPI pattern is strongly suppressed overall.
In contrast, noteworthy results are found for spin-orbit scattering,
\begin{align}
	\hat{V}_{\bk,\bkp} = V_0\{I+ic(\bk\times\bkp)\cdot\bm{\sigma}\},
	\label{eqn:SO}
\end{align}
where $c$ is the effective spin-orbit coupling parameter~\cite{Sinova15} and denotes strength of spin-orbit scattering relative to that of scalar scattering.
The $\Gbar$-$\Mbar$ peaks are selectively enhanced as $c$ increases and become stronger than the hexagonal ring as shown in Fig.~\ref{fig:calc}(c).

\begin{figure}
	\centering
	\includegraphics{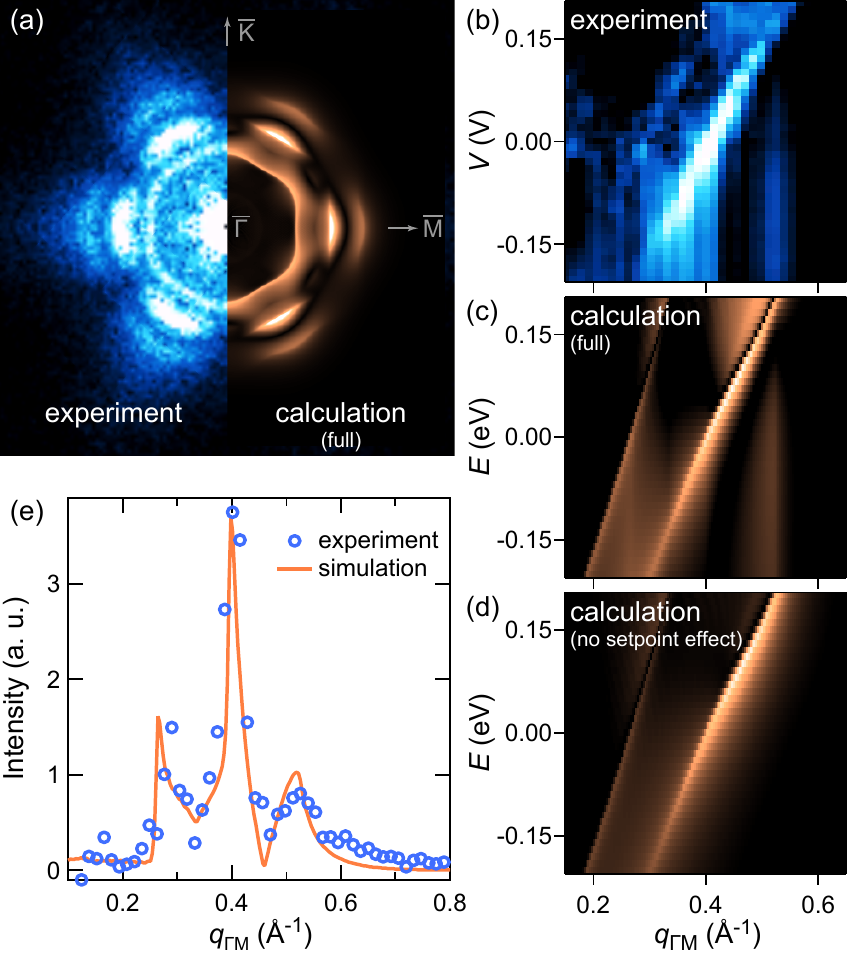}
	\caption{(color online) Full simulation including spin-orbit scattering with $c = 80$~\AA$^2$ and the setpoint effect with $V_\mathrm{set} = 0.2$~V\@.
	(a) A side-by-side comparison between the experiment (left) and the full simulation (right) at $-10$~meV\@.
	(b, c, d) Energy dependence (dispersion) in the $\Gbar$-$\Mbar$ direction.
	(e) Line profiles of (a) in the $\Gbar$-$\Mbar$ direction.
	An exponential background is subtracted from the experimental data.}
	\label{fig:comparison}
\end{figure}

The remaining feature, the humps in the $\Gbar$-$\Mbar$ direction, can be calculated by taking data acquisition procedures of SI-STM (the setpoint effect) into account.
Even if a $dI/dV$ spectrum is proportional to LDOS at each location as generally assumed, the proportional constant is not generally uniform but has a spatial structure reflecting variation of the tip height.
The height of a scanning tip is adjusted at each location such that the tunneling current is a set value.
The current is determined by the LDOS integrated up to a given bias voltage $V_\mathrm{set}$.
Consequently, a $dI/dV$ image observed by SI-STM depends on $V_\mathrm{set}$ as well as the LDOS,
\begin{align}
	\frac{dI}{dV}(\bm{r},V,V_\mathrm{set}) \propto \frac{\rho(\bm{r},eV)}{\displaystyle\int_{0}^{eV_\mathrm{set}}\!\!\rho(\bm{r},E_\mathrm{F}+\epsilon)d\epsilon},
	\label{eqn:dIdV}
\end{align}
where $E_\mathrm{F}$ is the Fermi energy~\cite{Kohsaka07}.
The denominator of Eq.~\eqref{eqn:dIdV} represents the setpoint effect.
This effect has been known in the experiments but neglected in the calculations of QPI.
Full simulation including spin-orbit scattering and the setpoint effect is shown in Fig.~\ref{fig:comparison}.
All of the hexagonal ring, the strong $\Gbar$-$\Mbar$ peaks, and the $\Gbar$-$\Mbar$ humps are well reproduced.
The peak intensities agree with the experiment as shown in Fig.~\ref{fig:comparison}(e) with $c = 80$~\AA$^2$ or a dimensionless parameter $c\pi^2/{a_0}^2=40$ ($a_0$ = 4.34~\AA, $a$-axis length), indicating predominance of spin-orbit scattering over scalar scattering.

\section{Discussion}

The above results of numerical simulations clearly show that spin-orbit scattering is the crucial ingredient to explain the QPI intensities of BiTeI\@.
The contrasting results of magnetic and spin-orbit scattering can be understood as follows.
Electrons with spin-up and spin-down feel potentials of opposite signs for magnetic scattering~\cite{Zhou09}.
In addition, electrons scattered to the right and the left do as well for spin-orbit scattering.
The opposite signs result in suppression of QPI for magnetic scattering whereas the same sign, as a result of a combination of the two effects, is cooperative for QPI in the case of spin-orbit scattering.
An essential point of this mechanism is the scattering amplitude.
To the first order of $\hat{V}$, a contribution to QPI from a scattering process ($\bkp\rightarrow\bk$) and its time-reversal counterpart ($-\bk\rightarrow-\bkp$) are written as
\begin{align}
	\delta\rho_{\bk,\bkp}(\omega)
	&= -\frac1\pi\sum_{m,n}\mathrm{Im}\big\{\gn\gm\big\}
		\big\langle\psin\big|\hat{V}_{\bk,\bkp}\big|\psim\big\rangle
		\big\langle\psim\big|\psin\big\rangle,&
		\label{eqn:rhokk}\\
	\delta\rho_{-\bkp,-\bk}(\omega)
	&= -\frac1\pi\sum_{m,n}\mathrm{Im}\big\{\gn\gm\big\}
		\big\langle\psin\big|\Theta\hat{V}_{-\bk,-\bkp}\Theta^{-1}\big|\psim\big\rangle
		\big\langle\psim\big|\psin\big\rangle,
	\label{eqn:rhokkTR}
\end{align}
where $\Theta$ is the time-reversal operator.
(Derivation of these formulas is written in Appendix~\ref{appendix:scatteringProcesses}.)
The difference between the two processes is found to be the potential in the scattering amplitude.
For scalar scattering $\hat{V}_{\bk,\bkp}=V_0I$, Eqs.~\eqref{eqn:rhokk} and \eqref{eqn:rhokkTR} are the same, $\delta\rho_{\bk,\bkp} = \delta\rho_{-\bkp,-\bk}$.
For magnetic scattering $\hat{V}_{\bk,\bkp}=V_0\sigma_i$, the scattering amplitude changes sign under time reversal because $\Theta\sigma_i\Theta^{-1}=-\sigma_i$.
Time-reversal processes thus always cancel with each other, $\delta\rho_{\bk,\bkp} = -\delta\rho_{-\bkp,-\bk}$, leading to the strong suppression of QPI for magnetic scattering.
(This explains why QPI patterns are unchanged even in the presence of magnetic impurities~\cite{Beidenkopf11,Strozecka11}.)
For spin-orbit scattering $\hat{V}_{\bk,\bkp} = icV_0(\bk\times\bkp)\cdot\bm{\sigma}$, the scattering amplitude of time-reversal processes has the same sign because $\Theta\hat{V}_{-\bk,-\bkp}\Theta^{-1}=(-i)cV_0(\bk\times\bkp)\cdot(-\bm{\sigma})=\hat{V}_{\bk,\bkp}$.
Therefore, as for spin-orbit scattering, the two scattering processes cooperatively contribute to QPI without being canceled as in the case of scalar scattering.

The preferential enhancement of the $\Gbar$-$\Mbar$ peaks (Fig.~\ref{fig:calc}(c)) is attributed to the directional and spin-dependent nature of spin-orbit scattering.
For electrons in two dimensions, spin-orbit scattering is written as $\hat{V}_{\bk,\bkp} = icV_0kk^\prime\sin\theta_{\bk,\bkp}\sigma_z$, where $\theta_{\bk,\bkp}$ is the angle from $\bk$ to $\bkp$.
Since the hexagonal ring mainly consists of $\delta\rho_{\bk,\bkp}$ with $\theta_{\bk,\bkp}\sim\pi$ and $\langle\psin|\sigma_z|\psim\rangle\sim 0$, spin-orbit scattering does not contribute to the hexagonal ring whereas it does contribute to the $\Gbar$-$\Mbar$ peaks because $\theta_{\bk,\bkp}\sim\pm2\pi/3$ and $\langle\psin|\sigma_z|\psi_n(\bkp)\rangle\neq 0$.
(Whether the scattering amplitude is zero or not is easily estimated as written in Appendix~\ref{appendix:amplitudeAndSpin}.)
This selective suppression is a selection rule originating from the scattering amplitude $\langle\psin|\hat{V}_{\bk,\bkp}|\psim\rangle = 0$, which is distinct from a selection rule stemming from orthogonal wave functions $\langle\psin|\psim\rangle = 0$~\cite{Petersen00,Pascual04,Brihuega08,Strozecka11,Beidenkopf11}.

Spin-orbit scattering exists in principle in any materials and grows with spin-orbit coupling.
However, its appearance in QPI depends on details of relevant electronic states.
QPI of surface states of Au(111)~\cite{Hasegawa93} is insensitive to spin-orbit scattering due to the selection rule, $\theta_{\bk,\bkp}\sim\pi$ and $\langle\psin|\sigma_z|\psim\rangle\sim 0$, being the same as the hexagonal ring of BiTeI\@.
As for topological surface states of Bi$_2$Te$_3$,~\cite{Zhang09,Beidenkopf11} we presume that spin-orbit scattering enhances the $\Gbar$-$\Mbar$ peaks so it does for BiTeI, but its influence remains to be clarified.
In this sense, the quasi-two-dimensional states of BiTeI with two scattering channels, one of which is sensitive to spin-orbit scattering and the other insensitive, are suited to investigate effects of spin-orbit scattering.
QPI arising from electronic states near the Brillouin-zone boundary may be subject to spin-orbit scattering because of large momenta and $\theta_{\bk,\bkp}\sim\pi/2$~\cite{Lee09}.
Such candidates are found in topological crystalline insulators and Weyl semimetals~\cite{Zeljkovic14,Zheng16,Inoue16}.

Including other factors affecting QPI intensities would be conducive to better quantifying strength of spin-orbit scattering.
A delta-function scattering potential is used for simplicity in our simulations.
In a more realistic case, $V_0$ in Eq.~\eqref{eqn:SO} is changed from a constant to $V(|\bm{q}|)$ for a spherical scattering potential.
Since $V(|\bm{q}|)$ usually decreases monotonically with increasing $|\bm{q}|$, QPI intensities are prone to be suppressed at large $|\bm{q}|$, where the amplitude of spin-orbit scattering is large.
Finite sharpness of a scanning tip also causes a similar effect.
Strength of spin-orbit scattering therefore may be underestimated due to these factors.
Nevertheless, the strength of spin-orbit scattering obtained for BiTeI (80~\AA$^2$) is much larger than a theoretical value for $n$-GaAs (5.3~\AA$^2$),~\cite{Sinova15} being consistent with strong spin-orbit coupling in BiTeI.
We note that the obtained value is averaged over many defects of multiple kinds.
If defects are separated enough, spin-orbit scattering can be probed at individual defects and may be available for designing and optimizing materials of spin Hall effect.

\section{conclusions}

In conclusion, using atomic-resolution SI-STM and numerical simulations with the $T$-matrix formalism, we identify a signature of spin-orbit scattering in the QPI patterns of BiTeI\@.
Spin-orbit scattering manifests itself in QPI through a selection rule originating from the scattering amplitude.
Our results highlight the importance of the scattering process beyond featureless scalar scattering and, more importantly, suggest a potential capability of QPI measurement as a local, direct (unaffected by scattering time), and quantitative probe of spin-orbit scattering detected heretofore by transport measurements.
We believe that including spin-orbit scattering into QPI analysis, which is now readily possible as demonstrated in our simulations, leads to a deeper understanding of electronic states of and more functionality from strong spin-orbit coupling systems.

\begin{acknowledgments}
We thank M. S. Bahramy and Wei-Cheng Lee for fruitful discussions.
This work is supported by the Murata Science Foundation.
\end{acknowledgments}

\appendix

\section{Fast calculation of QPI patterns\label{appendix:fastCalculation}}

The $\bk$ summation in Eq.~\eqref{eqn:rhoQ} with the integral equation about the $T$-matrix Eq.~\eqref{eqn:Tseries} requires $\mathcal{O}(N^4)$ operations for single $\bm{q}$, resulting in $\mathcal{O}(N^6)$ operations in total to calculate a QPI image.
In the case of momentum-independent scattering, the total amount of calculation is reduced to $\mathcal{O}(N^4)$ and further reduced to $\mathcal{O}(N^2\log_2N)$ by using FFT.
Here we show that, even in the case of momentum-dependent scattering, the amount of calculation can be reduced to $\mathcal{O}(N^2\log_2N)$ from $\mathcal{O}(N^6)$ of the $\bk$ summation when the scattering satisfies a certain condition.

To introduce our approach to reduce the amount of calculation, we begin with momentum-independent scattering.
Since the $T$ matrix is also independent of $\bk$, Eqs.\eqref{eqn:rhoQ}-\eqref{eqn:Tseries} are simplified to
\begin{align}
	\rho(\bm{q},\omega)
	\sim -\frac{1}{2\pi i}\sum_{\bk}\mathrm{Tr}\left\{\hat{G}_0(\bk,\omega)\hat{T}(\omega)\hat{G}_0(\bk-\bm{q},\omega)-\hat{G}_0^*(\bk,\omega)\hat{T}^*(\omega)\hat{G}_0^*(\bk+\bm{q},\omega)\right\},
	\label{eqn:simplerho}
\end{align}
where $\hat{T}(\omega)=\left(I-\hat{V}\sum_{\bk}\hat{G}_0(\bk,\omega)\right)^{-1}\hat{V}$.
Only the inhomogeneous part of LDOS is shown in Eq.~\eqref{eqn:simplerho} for brevity because it gives spatial modulations of QPI patterns.
Each matrix element of the $\bk$ summation in Eq.~\eqref{eqn:simplerho} is
\begin{align}
	\sum_{\bk}\mathrm{Tr}\left\{\hat{G}_0(\bk,\omega)\hat{T}(\omega)\hat{G}_0(\bk+\bm{q},\omega)\right\}
	= \sum_{m,n}t^{(mn)}(\omega)\sum_{\bk,j}g_0^{(jm)}(\bk,\omega)g_0^{(nj)}(\bk+\bm{q},\omega),
	\label{eqn:matrixElement}
\end{align}
where $g_0^{(ij)}$ and $t^{(ij)}$ are matrix elements of $\hat{G}_0$ and $\hat{T}$, respectively.
The right-hand side of Eq.~\eqref{eqn:matrixElement} is the cross-correlation between $g_0^{(jm)}$ and $g_0^{(nj)}$, and thus can be expressed with Fourier transform,
\begin{align}
	\sum_{\bk}g_0^{(jm)}(\bk,\omega)g_0^{(nj)}(\bk+\bm{q},\omega)
	= \sum_{\bm r}g_0^{(jm)}(-\bm{r},\omega)g_0^{(nj)}(\bm{r},\omega)e^{i\bm{q}\cdot\bm{r}},
	\label{eqn:CrossCorrelation}
\end{align}
where $g_0^{(ij)}(\bm{r},\omega)=\sum_{\bk}g_0^{(ij)}(\bk,\omega)e^{-i\bk\cdot\bm{r}}$.
Since the right-hand side of Eq.~\eqref{eqn:CrossCorrelation} is the inverse Fourier transform of $g_0^{(jm)}(-\bm{r},\omega)g_0^{(nj)}(\bm{r},\omega)$, the left-hand side of Eq.~\eqref{eqn:CrossCorrelation} can be calculated via FFT without taking the $\bk$ summation~\cite{Arguello15,Derry15}.
The amount of calculation is thus reduced from $\mathcal{O}(N^4)$ of the direct calculation to $\mathcal{O}(N^2\log_2N)$ of FFT\@.

FFT, the essential point to reduce the amount of calculation, is available as in Eq.~\eqref{eqn:CrossCorrelation} because Eq.~\eqref{eqn:simplerho} is expressed virtually as a product of two matrices; one is a function of $\bm{k}$ and the other is a function of $\bm{k}+\bm{q}$.
At a glance, this condition is not satisfied for general scatterers because the $T$ matrix depends on momentum.
However, when $\hat{V}_{\bm{k},\bm{k}^\prime}$ is expressed as a sum of products between $\bm{k}$- and $\bm{k}^\prime$-dependent matrices
\begin{align}
	\hat{V}_{\bk,\bkp}
	= \sum_j\hat{u}_j(\bk)\hat{v}_j(\bkp)
	= {}^t\!\bm{u}(\bk)\bm{v}(\bkp),
	\label{eqn:V}
\end{align}
where $\bm{u}(\bk)={}^t\!\left(\hat{u}_1(\bk)\ \hat{u}_2(\bk)\ \cdots\right)$ and $\bm{v}(\bk)={}^t\!\left(\hat{v}_1(\bk)\ \hat{v}_2(\bk)\ \cdots\right)$, FFT is available to calculate QPI patterns as in the case of momentum-independent scattering.

We rewrite Eq.~\eqref{eqn:Tseries} in a form of a recurrence formula
\begin{align}
	\hat{T}^{(n)}_{\bk,\bkp}(\omega) =
	\begin{cases}
		\hat{V}_{\bk,\bkp} & (n = 1),\\
		\hat{V}_{\bk,\bkp} + \sum_{\bm p}\hat{V}_{\bk,\bm{p}}\hat{G}_0(\bm{p},\omega)\hat{T}^{(n-1)}_{\bm{p},\bkp}(\omega) & (n \geq 2).
	\end{cases}
	\label{eqn:Tn}
\end{align}
By multiplying $\bm{v}(\bk)\hat{G}_0(\bk,\omega)$ from the left and taking a sum with respect to $\bk$, we obtain
\begin{align}
	K_n(\bkp,\omega) =
	\begin{cases}
		M(\omega)\bm{v}(\bkp) & (n = 1),\\
		M(\omega)\bm{v}(\bkp) + M(\omega)K_{n-1}(\bkp,\omega) & (n \geq 2),
	\end{cases}
\end{align}
where $K_n(\bkp,\omega)=\sum_{\bm p}\bm{v}(\bm{p})\hat{G}_0(\bm{p},\omega)\hat{T}^{(n)}_{\bm{p},\bkp}(\omega)$ and $M(\omega) = \sum_{\bm p}\bm{v}(\bm{p})\hat{G}_0(\bm{p},\omega){}^t\!\bm{u}(\bm{p})$.
Since these equations are summarized as $K_n(\bkp,\omega) = \sum_{j=1}^n\{M(\omega)\}^j\bm{v}(\bkp)$, Eq.~\eqref{eqn:Tn} is also summarized as
\begin{align}
	\hat{T}^{(n)}_{\bk,\bkp}(\omega) = {}^t\!\bm{u}(\bk)\sum_{j=0}^{n-1}\{M(\omega)\}^j\bm{v}(\bkp),
	\label{eqn:TnFinal}
\end{align}
where $\{M(\omega)\}^0$ is the identity matrix, ${}^t\!\bm{u}(\bk)\{M(\omega)\}^0\bm{v}(\bkp) = {}^t\!\bm{u}(\bk)\bm{v}(\bkp)$.
Now the $T$ matrix and thus the second term of Eq.~\eqref{eqn:G} are expressed as a product of $\bk$  and $\bkp$ terms, and FFT is available to calculate QPI patterns as described above.
We stress that the integral equation of the $T$ matrix [Eq.~\eqref{eqn:Tseries}] is reduced virtually to a sum of $\{M(\omega)\}^j$, which is independent of $\bk$.
Therefore, the calculation size is almost the same as that of momentum-independent scatterers.
Namely, the total amount of QPI calculation is drastically reduced from $\mathcal{O}(N^6)$ to $\mathcal{O}(N^2\log_2N)$.

\begin{figure}
	\centering
	\includegraphics{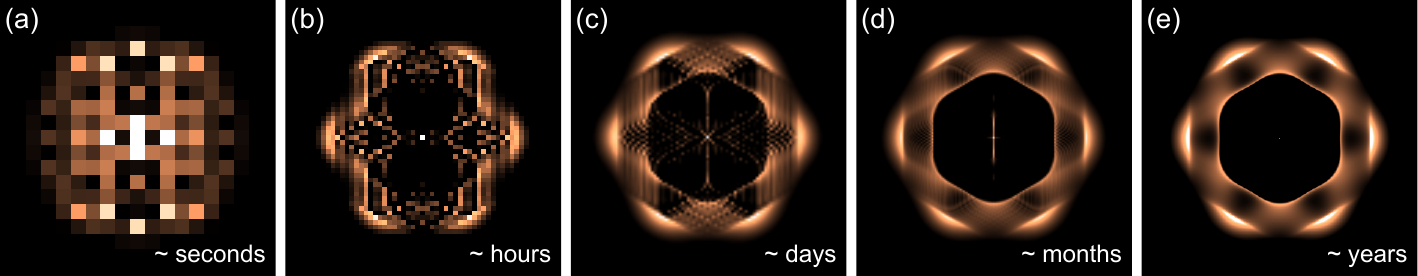}
	\caption{(color online) Simulations with the same parameters as Fig.~\ref{fig:calc}(c) but at lower resolutions.
		Since the calculation of Fig.~\ref{fig:calc}(c) took 4 s, calculations with direct $\bk$ summation of (a)--(e) are estimated to take 4 s, 1 h, 1 day, 1 month, and 7 years, respectively.
		(e) The same as Fig.~\ref{fig:calc}(c) shown for comparison.}
	\label{fig:lowResSim}
\end{figure}

Spin-orbit scattering, $\hat{V}_{\bk,\bkp} = V_0\left\{I + ic(\bk\times\bkp)\cdot\bm{\sigma}\right\}$, is written in the form of Eq.~\eqref{eqn:V}.
In two dimension, it is written as $\hat{V}_{\bk,\bkp} = V_0\left\{I + ic(k_xk_y^\prime-k_yk_x^\prime)\sigma_z\right\} = {}^t\!\bm{u}(\bk)\bm{v}(\bkp)$, where ${}^t\!\bm{u}(\bk)=(I\ k_xI\ k_yI)$, $\bm{v}(\bk)=U\bm{u}(\bk)$, and 
\begin{align}
	U = \begin{pmatrix}
		V_0I	& 0					& 0 \\
		0		& 0					& icV_0\sigma_z \\
		0		& -icV_0\sigma_z	& 0
	\end{pmatrix}.
\end{align}
The $M$ matrix is
\begin{align}
	M(\omega)
	&= U\sum_{\bk}\begin{pmatrix}
		\hat{G}_0(\bk,\omega)	& k_x\hat{G}_0(\bk,\omega)		& k_y\hat{G}_0(\bk,\omega)\\
		k_x\hat{G}_0(\bk,\omega)& k_x^2\hat{G}_0(\bk,\omega)	& k_xk_y\hat{G}_0(\bk,\omega)\\
		k_y\hat{G}_0(\bk,\omega)& k_xk_y\hat{G}_0(\bk,\omega)	& k_y^2\hat{G}_0(\bk,\omega)
	\end{pmatrix}.
\end{align}
Note that each element of $M$ is a $2\times 2$ matrix independent of $\bm{k}$ and thus $\sum M^j$ is readily calculable.
Calculation of Fig.~\ref{fig:calc}(c) with this method took only several seconds with a desktop computer.
This means direct $\bk$ summation at the same resolution is estimated to take about several years.
Even if the direct calculations were done at lower resolutions, it may be difficult to find a reasonable compromise between resolution and time as shown in Fig.~\ref{fig:lowResSim}, highlighting drastic reduction of calculation costs with our method.

\section{Limitations of the joint density of states approach\label{appendix:JDOS}}

All QPI patterns observed in the experiments show large intensities near $\bm{q}=0$ as shown in Fig.~\ref{fig:exp}(b).
Since these near-$\Gbar$ features vary from one field of view to another, and one sample to another, they originate from the nanoscale inhomogeneity due to random distribution of defects.
The varying near-$\Gbar$ feature means that a QPI pattern near $\bm{q}=0$, if any, is small and masked by the nanoscale inhomogeneity.
Therefore, no prominent feature near $\bm{q}=0$ in Fig.~\ref{fig:calc} is consistent with the experiments.

Meanwhile, from the viewpoint of the so-called joint density of states (JDOS), one may expect a large QPI intensity near $\bm{q}=0$;
if DOS is large at a given $\bk$, large DOS is also found near $\bk$, resulting in a large QPI intensity near $\bm{q}=0$.
However, as revealed by JDOS calculations shown below, such a JDOS-derived pattern near $\bm{q}=0$ has never been observed in BiTeI.

\begin{figure}
	\centering
	\includegraphics{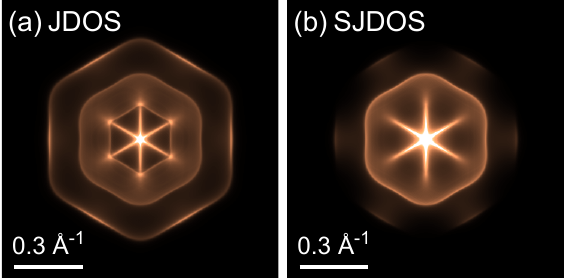}
	\caption{(color online) Fourier-transformed images of QPI patterns at $-10$~meV, calculated by (a) JDOS and (b) spin-dependent JDOS.}
	\label{fig:JDOS}
\end{figure}

In the JDOS approach, QPI patterns are interpreted to be proportional to JDOS,
\begin{align}
	\rho^\mathrm{JDOS}(\bm{q},\omega) \propto \sum_{\bk}\rho_0(\bk,\omega)\rho_0(\bk-\bm{q},\omega),
	\label{eqn:JDOS}
\end{align}
where $\rho_0(\bk,\omega)$ is the DOS at $\bk$, $\rho_0(\bk,\omega)=-\mathrm{Im}\Big[\mathrm{Tr}\big\{G_0(\bk,\omega)\big\}\Big]$.
To include a spin effect, spin-dependent JDOS is considered,
\begin{align}
	\rho^\mathrm{SJDOS}(\bm{q},\omega) \propto 
	\sum_{i=0,x,y,z}\sum_{\bk}\rho_i(\bk,\omega)\rho_i(\bk-\bm{q},\omega)
	\label{eqn:SJDOS}
\end{align}
where $\rho_i(\bk,\omega)=-\mathrm{Im}\Big[\mathrm{Tr}\big\{\sigma_iG_0(\bk,\omega)\big\}\Big]$.
The JDOS approach always predicts a large QPI intensity near $\bm{q}=0$ because the JDOS is the auto-correlation of $\rho_i(\bk,\omega)$.
Actually, as shown in Fig.~\ref{fig:JDOS}, calculations with the JDOS and spin-dependent JDOS approaches show an asterisk-like pattern centered at $\bm{q}=0$, sticking out in the $\Gbar$-$\Kbar$ direction, and extending close to the hexagonal ring of intraband scattering.
However, such a salient pattern has never been observed in the experiments.
This discrepancy between the JDOS calculations and the observed QPI patterns highlights limitations of the JDOS approach.

The limitations of the JDOS approach derive from a difference between QPI and the JDOS.
The spin-dependent JDOS of Eq.~\eqref{eqn:SJDOS} can be written as
\begin{align}
	\rho^\mathrm{SJDOS}(\bm{q},\omega) \propto 
	\sum_{\bk}\sum_{m,n}\mathrm{Im}\big\{\gn\big\}\mathrm{Im}\big\{g_m(\bk-\bm{q},\omega)\big\}
	\Big|\big\langle\psi_{m}(\bk-\bm{q})\big|\psin\big\rangle\Big|^2.
	\label{eqn:SJDOS2}
\end{align}
A QPI pattern calculated by the $T$-matrix formalism for scalar scattering is written in a similar form.
Equation~\eqref{eqn:rhokk} with $\hat{V}_{\bk,\bkp}=V_0I$ gives
\begin{align}
	\rho(\bm{q},\omega) = \sum_{\bk}\delta\rho_{\bk,\bk-\bm{q}}(\omega) \propto 
	\sum_{\bk}\sum_{m,n}\mathrm{Im}\big\{\gn g_m(\bk-\bm{q},\omega)\big\}
	\Big|\big\langle\psi_{m}(\bk-\bm{q})\big|\psin\big\rangle\Big|^2.
	\label{eqn:rhokkScalar}
\end{align}
The product of $\mathrm{Im}\,g$ in Eq.~\eqref{eqn:SJDOS2} is not included in Eq.~\eqref{eqn:rhokkScalar}; in other words, that QPI is not directly related to JDOS as discussed in Ref.~\onlinecite{Derry15}.

\section{Contribution to QPI from $\bk\rightarrow\bkp$ scattering process\label{appendix:scatteringProcesses}}

We define $\delta\rho_{\bk,\bkp}(\omega)$ such that its summation with respect to $\bk$ gives $\rho(\bk-\bkp,\omega)$,
\begin{align}
	\delta\rho_{\bk,\bkp}(\omega) \equiv -\frac{1}{2\pi i}\mathrm{Tr}\left\{\hat{G}(\bk,\bkp,\omega)-\hat{G}^*(\bkp,\bk,\omega)\right\}.
\end{align}
Here we consider an approximation to the first order of $\hat{V}_{\bk,\bkp}$ for simplicity.
Since the $T$ matrix is $\hat{T}_{\bk,\bkp} \sim \hat{V}_{\bk,\bkp}$,
\begin{align}
	\delta\rho_{\bk,\bkp}(\omega) 
	= -\frac{1}{2\pi i}\mathrm{Tr}\left[\hat{G}_0(\bk,\omega)\hat{V}_{\bk,\bkp}\hat{G}_0(\bkp,\omega)-\left\{\hat{G}_0(\bkp,\omega)\hat{V}_{\bkp,\bk}\hat{G}_0(\bk,\omega)\right\}^*\right].
\end{align}
The first term is
\begin{align}
	& \mathrm{Tr}\left\{\hat{G}_0(\bk,\omega)\hat{V}_{\bk,\bkp}\hat{G}_0(\bkp,\omega)\right\}\nonumber\\
	& = \sum_{m,n}\gn\gm\mathrm{Tr}\left\{\big|\psin\big\rangle\big\langle\psin\big|\hat{V}_{\bk,\bkp}\big|\psim\big\rangle\big\langle\psim\big|\right\}\\
	& = \sum_{m,n}\gn\gm\big\langle\psin\big|\hat{V}_{\bk,\bkp}\big|\psim\big\rangle\big\langle\psim\big|\psin\big\rangle.
	\label{eqn:firstTerm}
\end{align}
Similarly, the second term is
\begin{align}
	\mathrm{Tr}\left\{\hat{G}_0(\bkp,\omega)\hat{V}_{\bkp,\bk}\hat{G}_0(\bk,\omega)\right\}^*
	= \sum_{m,n}\big\{\gn\gm\big\}^*\big\langle\psin\big|\hat{V}_{\bkp,\bk}{}^\dagger\big|\psim\big\rangle\big\langle\psim\big|\psin\big\rangle.
	\label{eqn:secondTerm}
\end{align}
Given $\hat{V}_{\bkp,\bk}{}^\dagger = \hat{V}_{\bk,\bkp}$ as $\hat{V}$ is Hermitian, Eqs.~\eqref{eqn:firstTerm} and \eqref{eqn:secondTerm} are summarized to
\begin{align}
	\delta\rho_{\bk,\bkp}(\omega)
	= -\frac1\pi\sum_{m,n}\mathrm{Im}\big\{\gn\gm\big\}
		\big\langle\psin\big|\hat{V}_{\bk,\bkp}\big|\psim\big\rangle
		\big\langle\psim\big|\psin\big\rangle.
		\tag{\ref{eqn:rhokk}}
\end{align}
Replacing $\bk$ ($\bkp$) with $-\bkp$ ($-\bk$) gives the time-reversal counterpart,
\begin{align}
	&\delta\rho_{-\bkp,-\bk}(\omega)\nonumber\\
	&= -\frac1\pi\sum_{m,n}\mathrm{Im}\big\{g_n(-\bk,\omega)g_m(-\bkp,\omega)\big\}
		\big\langle\psi_m(-\bkp)\big|\hat{V}_{-\bkp,-\bk}\big|\psi_n(-\bk)\big\rangle
		\big\langle\psi_n(-\bk)\big|\psi_m(-\bkp)\big\rangle\\
	&= -\frac1\pi\sum_{m,n}\mathrm{Im}\big\{\gn\gm\big\}
		\big\langle\psin\big|\Theta\hat{V}_{-\bkp,-\bk}{}^\dagger\Theta^{-1}\big|\psim\big\rangle
		\big\langle\psim\big|\psin\big\rangle.
\end{align}
The second follows from three identities;
$g_n(-\bk,\omega)=\gn$ as $H_0$ is time invariant, and 
$\big\langle\beta\big|{\cal L}\big|\alpha\big\rangle=\big\langle\alpha^\prime\big|\Theta {\cal L}^\dagger\Theta^{-1}\big|\beta^\prime\big\rangle$ and $\big\langle\beta\big|\alpha\big\rangle = \big\langle\alpha^\prime\big|\beta^\prime\big\rangle$, where $\big|\alpha^\prime\big\rangle=\Theta\big|\alpha\big\rangle$, $\big|\beta^\prime\big\rangle=\Theta\big|\beta\big\rangle$, and $\cal L$ is a linear operator~\cite{JJSakurai}.
The last two hold because $\Theta$ is antiunitary.
Given $\hat{V}_{-\bkp,-\bk}{}^\dagger = \hat{V}_{-\bk,-\bkp}$ as $\hat{V}$ is Hermitian, we obtain
\begin{align}
	\delta\rho_{-\bkp,-\bk}(\omega)
	&= -\frac1\pi\sum_{m,n}\mathrm{Im}\big\{\gn\gm\big\}
		\big\langle\psin\big|\Theta\hat{V}_{-\bk,-\bkp}\Theta^{-1}\big|\psim\big\rangle
		\big\langle\psim\big|\psin\big\rangle.
	\tag{\ref{eqn:rhokkTR}}
\end{align}
%

\section{Relation between the scattering amplitude and spin orientation\label{appendix:amplitudeAndSpin}}

Let $\bm{S}_n(\bk)$ be the expectation value of the Pauli matrices, $\bm{S}_n(\bk) = \big\langle\psin\big|\bm{\sigma}\big|\psin\big\rangle$, then the following relation holds
\begin{align}
	\bm{S}_n(\bk)\cdot\bm{S}_m(\bkp) = 2\Big|\big\langle\psin\big|\psim\big\rangle\Big|^2-1.
\end{align}
The $\pi$-rotation operator along the $z$ direction for a spin-1/2 state is given as $e^{-i\pi\sigma_z/2}=-i\sigma_z$.
Therefore, by rotating $\bm{S}_m(\bkp)$ and $\big|\psim\big\rangle$ by $\pi$ along the $z$ direction, we obtain
\begin{align}
	\bm{S}_n(\bk)\cdot R_{z\pi}\bm{S}_m(\bkp) = 2\Big|\big\langle\psin\big|\sigma_z\big|\psim\big\rangle\Big|^2-1,
	\label{eqn:SdotRS}
\end{align}
where $R_{z\pi}$ denotes $\pi$-rotation of spin orientation along the $z$ direction.
Equation~\eqref{eqn:SdotRS} means $\big\langle\psin\big|\sigma_z\big|\psim\big\rangle = 0$ for two states with the spin orientations parallel and lying in the $xy$ plane.

\input{SOscattering.bbl}

\end{document}

%% file: SOscattering.bbl
%